\documentclass[twocolumn]{aastex63}

\usepackage{graphicx}   
\hypersetup{linkcolor=red,citecolor=green,filecolor=cyan,urlcolor=magenta}

\newcommand{\cii}{\mbox{[CII]}}
\newcommand{\co}{\mbox{$^{12}$CO}}

\newcommand{\kms}{\mbox{km s$^{-1}$}}
\def \HI{H\,{\sc i}}
\newcommand{\htwo}{\mbox{H$_2$}}

\newcommand{\cc}{\mbox{cm$^{-3}$}}
\newcommand{\cmsq}{\mbox{cm$^{-2}$}}
\newcommand{\msun}{\mbox{M$_\odot$}}

\newcommand{\vlsr}{\mbox{$V_{{\rm LSR}}$}}

\newcommand{\etal}{\mbox{et~al.~}}

\received{}
\revised{}
\accepted{}

\shorttitle{CII, CO, \HI~21~cm}
\shortauthors{Heyer \etal }

\begin{document}

\title{Searching for converging flows of atomic gas onto a molecular cloud}  

\correspondingauthor{Mark Heyer}
\email{heyer@astro.umass.edu}

\author[0000-0002-3871-010X]{Mark Heyer}
\affiliation{Astronomy Department, University of Massachusetts,
Amherst, MA, 01003 USA}
\author[0000-0002-6622-8396]{Paul F. Goldsmith}
\affiliation{Jet Propulsion Laboratory, California Institute of Technology, 
4800 Oak Grove Drive, Pasadena, CA 91109, USA}
\author[0000-0003-2555-4408]{Robert Simon}
\affiliation{I. Physikalisches Institut der Universit{\"a}t zu K{\"o}ln,
Z{\"u}lpicher Sta{\ss}e 77, 50937, K{\"o}ln, Germany}
\author[0000-0002-1316-1343]{Rebeca Aladro}
\affiliation{Max-Planck-Institut f{\"u}r RadioAstronomie, Auf dem H{\"u}gel 69, D-53121, 
Bonn, Germany}
\author{Oliver Ricken}
\affiliation{Max-Planck-Institut f{\"u}r RadioAstronomie, Auf dem H{\"u}gel 69, D-53121, 
Bonn, Germany}


\begin{abstract}
We present new observations of \cii\ $^2P_{3/2} \rightarrow ^2P_{1/2}$ fine structure line emission from an isolated 
molecular cloud using the upGREAT instrument onboard {\it SOFIA}.  
These data are analyzed together with archival CO $J$=1--0 and \HI~21~cm emission spectra to investigate the role of converging atomic gas flows in the formation of molecular clouds.  
Bright \cii\ emission is detected throughout the mapped area that likely originates from photodissociation regions excited by UV radiation 
fields produced by newborn stars within the cloud.  
Upon spatial averaging of the \cii\ spectra, we identify weak \cii\ emission within velocity intervals where the \HI~21~cm line is brightest; these are blueshifted relative to velocities of the CO and bright \cii\ emission by 4~\kms.  
The brightness temperatures, velocity dispersions, and alignment with \HI~21~cm velocities connect this  \cii\ emission component 
to the cold, neutral atomic gas of the interstellar medium (CNM).  
We propose that this CNM feature is an accretion flow onto the far--side of the existing molecular  cloud.  
The mass infall rate is 3.2$\times$10$^{-4}$ {\msun}yr$^{-1}$.  
There is no direct evidence of a comparable red--shifted component in the \cii\ or HI~21~cm spectral lines that would indicate the presence of a converging flow.
\end{abstract}

\keywords{Interstellar clouds: molecular clouds -- Interstellar clouds: Photodissociation regions -- Interstellar clouds: Neutral atomic hydrogen clouds 
Interstellar phases: Molecular gas -- Interstellar phases: Cold neutral medium}

\section{Introduction} \label{sec:intro}
A key step in the star formation sequence is the assembly of interstellar clouds of near--fully molecular gas in which
conditions are most favorable for generating new stars.
This transition from atomic to molecular gas is fundamentally a microscopic, chemical process in which \htwo\ 
molecules form on grain surfaces while other molecular constituents develop via gas--phase or gas--grain chemistry.
However, macroscopic interstellar mechanisms are also  required to accumulate gas and dust into more opaque, high density, discrete configurations that allow these chemical reactions to take place and more importantly, to sustain high molecular gas fractions.
Specifically, sufficient column densities of dust grains and \htwo\ are required to self--shield \htwo\ molecules (and CO, the primary gas coolant capable of reducing gas temperatures below $\simeq$ 100~K and the spectroscopic tracer of cold \htwo) from dissociating far ultraviolet radiation \citep{Hollenbach:1971}.
Over the last 50 years, there have been many proposed macroscopic mechanisms to form molecular clouds.
These include gravitational instabilities in spiral arms and the agglomeration of existing, small molecular clouds into
giant molecular clouds \citep{McKee:2007, Dobbs:2014}.
To date, there has been no observational validation of any mechanism.

A recent development based on an earlier concept of cloud assembly posits that molecular clouds emerge within the
interface regions of large scale converging streams of neutral atomic gas  that develop from expanding supernova remnants, stellar winds, or
spiral density waves \citep{Hennebelle:1999, Ballesteros:1999, Heitsch:2006, Vazquez:2009}.  
The interaction of two converging streams produces shocks and triggers both gravitational and thermal instabilities that drive 
more gas from the warm, neutral medium (WNM) phase into the cold, neutral medium (CNM) phase, and ultimately, into the molecular gas phase.  
The resultant high gas volume density of the interface regions enables rapid formation of \htwo\ on dust grains.
Once established, a molecular cloud can continue to accrete material from these extended flows, which may sustain star formation activity 
over several 
cloud free--fall time scales.  

While the \HI~21~cm line would appear to be the obvious probe of an atomic gas flow,
\HI~21~cm emission does not distinguish between the more widely distributed warm neutral gas and the cold, neutral component that is the likely atomic precursor to molecular gas.
In addition, the rotational motions of the Galaxy, as well as non--circular motions, conspire to blend the \HI\ emission from physically unrelated volumes
of gas, making the spatial and kinematic connection of the atomic gas to the underlying molecular cloud in CO emission very uncertain.

Computational simulations of converging flows with simplified chemical networks to track carbon--bearing constituents and radiative transfer to determine the excitation of different energy levels  identify the  $^2P_{3/2} \rightarrow ^2P_{1/2}$ fine structure line of ionized carbon, \cii, at a wavelength of 158\micron, as a key tracer of the  cloud assembly process \citep{Franeck:2018, Clark:2019}.
Specifically, in the conditions in which the H to \htwo\ transition occurs, \cii\  emission  can trace the distribution and kinematics of the cold, neutral atomic gas and the layer of gas where hydrogen is mostly in the \htwo\ phase but the abundance of CO is low -- making its rotational lines very faint or undetectable (hereafter dark CO gas).

\citet{Clark:2019} identify two kinematic features in the \cii\ components that indicate the presence of a converging flow.  
These features depend on the initial conditions  of the flows -- specifically, the approaching head--on velocities, which they set at $\pm$3.75~\kms\
corresponding to a combined collisional velocity of 7.5~\kms.
First, they predict faint \cii\ emission from the cold, atomic neutral gas that carries the velocity information about the converging flow. 
The velocity offset of this emission is predicted to be 2--4 \kms\ from the velocity of CO emission seen in their simulation.  
Second, like the CNM gas, the CO dark component of \cii\ emission is spatially distributed outside of the more strongly self--shielded CO--emitting cloud, but its velocity corresponds more closely with the velocity interval of the CO emission.

In this investigation, we examine the role of converging atomic flows in the formation of  the compact molecular cloud BKP~7323 \citep{Brunt:2003} 
with new imaging observations of \cii\ 158\micron\ and archival data of CO J=1--0 line emission, and \HI~21~cm emission.
At a Galactic latitude of 4.24$^\circ$ and located beyond the Solar circle, BKP~7323 is well isolated both spatially and kinematically from any foreground and background molecular clouds.
This isolation facilitates the identification of large scale atomic gas flows associated with the  cloud that may be present. 
\citet{Brunt:2003} estimate the molecular mass of the cloud to be 1100~\msun, assuming a CO--to--\htwo\ conversion factor of 4.3 \msun/(K\,km\,s$^{-1}$\,pc$^2$) \citep{Bolatto:2013}  and a distance of 3.23~kpc \citep{Ragan:2012}.  
Under the 
designation ISOSJ22478+6357, 
this region has been the target of several previous studies investigating the early stages of star formation activity 
\citep{Hennemann:2008, Ragan:2012, Beuther:2021}. 

\section{Data} \label{sec:data}
\subsection{\cii\ Emission} 
\cii\ observations were obtained on the dates 23,25 February and 5 March 2021 as part of Stratospheric Observatory for Infrared Astronomy (SOFIA) Cycle 8 program 08-0062 
using the dual polarization, double sideband, 2$\times$7 pixel upGREAT Low Frequency Array (LFA)  that operates between 1.8 and 2.5~THz \citep{Risacher:2016}.  
The backends were fast Fourier transform spectrometers developed at the Max-Planck-Institut f{\"u}r  Radio Astronomie in Bonn \citep{Klein:2012}.  
These provide 4~GHz of bandwidth per pixel
and a maximum of 142~kHz spectral resolution corresponding to 630~\kms\ and 0.022~\kms\ velocity coverage and resolution respectively at the frequency 
of the \cii\ $^2P_{3/2} \rightarrow ^2P_{1/2}$ fine structure line.
Data were calibrated approximately every 5 minutes using measurements of the sky, and hot and cold loads.
Atmospheric transmission is estimated from the fits of atmospheric model brightness to the 
observed, calibrated sky-hot spectrum \citep{Guan:2012}.
Calibration uncertainties are estimated to be 20\% with the atmospheric transmission uncertainty contributing the largest error of 15\%.
\citep{Risacher:2016}.   Telluric lines  were present in our spectra but at velocities far displaced from our target and did not impact our science 
analysis and interpretation. 
At the wavelength of the \cii\ line, 
the half power beam width of our data is 14.1\arcsec.  Pointing and focus observations are taken at the beginning of each flight series.  During the 
flight, optical star pointing with three guide cameras are used to maintain the stellar pattern in their respective fields of view.  This procedure  provides
telescope pointing and tracking accuracy of 1\arcsec.

On--the--Fly (OTF) scans  were made along the Galactic longitude and latitude axes to form a cross pattern with overlap in coverage in the central region. 
The array was rotated by 19.1$^\circ$ with respect to the scanning direction to provide equal spacing along an OTF scan.  A second scan was executed in the reverse 
direction following a shift of 5.5\arcsec\ perpendicular to the scanning direction to achieve full sampling. 
The antenna scanning speed was 5.5\arcsec/sec and each scan length 
was 126.5\arcsec.  These scans were repeated multiple times (44 to 72) to improve the sensitivity. 

The data were processed with the  GREAT pipeline at the Max-Planck Institute for RadioAstronomy.  This included first order baseline subtraction and coadding spectra.
All coadded spectra were smoothed  to a spectral resolution of 0.25~\kms. 
Antenna temperatures were corrected for main beam efficiencies that range from 0.63 to 0.69 for the upGREAT/LFA beams placing the spectra on the mean beam 
temperature scale. 
The median rms per spectral channel over the coadded map is 0.22~K in main beam tempearture units.  

\subsection{Molecular Line Emission} \label{subsec:fcrao}
\co\ J=1--0 data for BKP~7323 were extracted from the FCRAO survey of the outer Galaxy \citep{Heyer:1998}.  
The angular resolution of the survey data,  sampled every 50\arcsec,  is 45\arcsec, with spectral resolution 0.81~\kms\ per channel.  
The median sensitivity per channel is 1~K in main beam temperature units.   
For this study, we smoothed the data to 60\arcsec\ angular resolution and 1~\kms\ spectral channels, which provides a median rms of 0.63~K in main beam temperature units. 

\subsection{HI 21~cm Line Emission} \label{subsec:drao}
Spectroscopic data cubes of \HI~21~cm line emission from the Canadian Galactic Plane Survey (CGPS) collected by the Dominion Radio Astronomical Observatory (DRAO) \citep{Taylor:2003} were obtained from the Canadian Astronomical Data Centre (CADC).  
The angular and spectral resolutions of these data are 1\arcmin\ and 0.82~\kms, respectively.  
The rms noise in brightness temperature is 3 to 4 K  per spectral channel.

\section{Results} \label{sec:results}
\begin{figure*}
\centering
\includegraphics[width=0.7\textwidth]{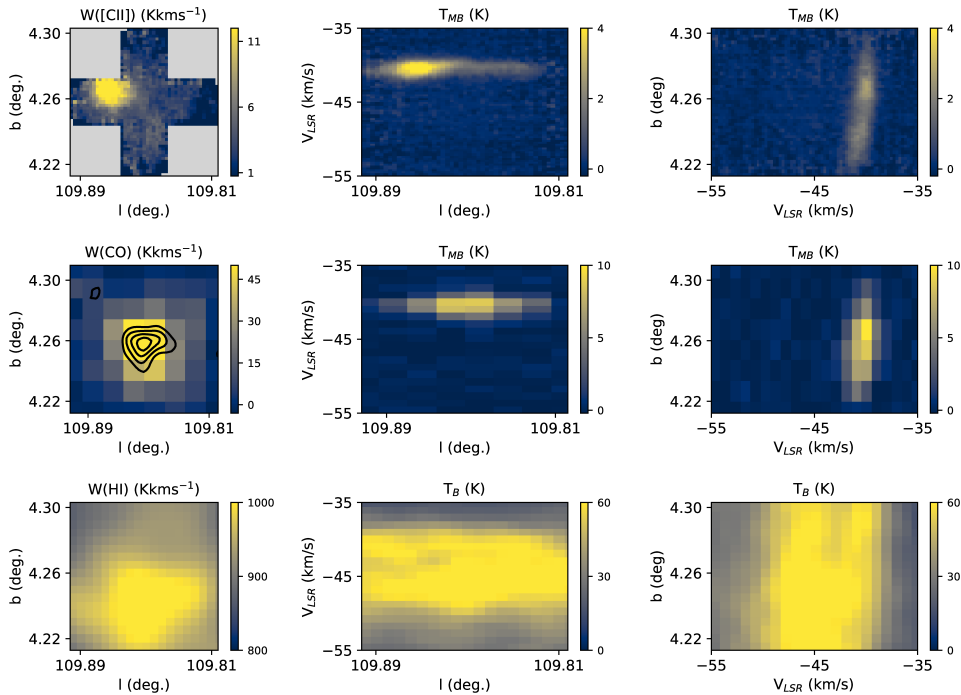}
\caption{Images of \cii, \co, and HI~21~cm\ line emission.  
From left to right: integrated intensity; longitude--velocity image; latitude--velocity image. 
The longitude-velocity images show spectra averaged over Galactic latitudes 4.213$^\circ$ to 4.303$^\circ$ and 
the latitude-velocity images show spectra averaged over Galactic longitudes 109.806$^\circ$ to 109.900$^\circ$.
(top row) representations of \cii\ emission.  The scaling of the \cii\ integrated intensity image begins at 0.75 K \kms\ that corresponds to 3$\sigma$ uncertainty.
Pixels with values below this threshold are dark and uniform.
(middle row) representations of \co\ emission.  The contours show the SCUBA 850\micron\ emission observed by 
\citet{DiFrancesco:2008}.  The contour levels are 0.15, 0.25, 0.35, 0.45 Jy/beam. 
(bottom row) representations of \HI~21~cm emission.
\label{fig:fig1}}
\end{figure*}

The spectral line data presented in this study provide information on the distribution and kinematics of both molecular 
and atomic interstellar gas. 
Figure~\ref{fig:fig1} shows three representations of the data for each spectral line.
From left to right we display the velocity--integrated emission, the Galactic longitude--velocity image with spectra averaged 
over the Galactic latitude range covered, 
and the Galactic latitude--velocity image with spectra averaged over the Galactic longitude range covered.
For \cii\ and \co, the velocity interval of the integration is $-43$ to $-38$~\kms.  For the \HI~21~cm data, the integration interval is $-55$ to $-35$~\kms. 

The gas column density distributions are encapsulated in the integrated intensity images of \cii, CO, and HI~21cm. 
\cii\ emission is widespread throughout the cross area of the map with 79\% of the observed \cii\ spectra detected with integrated intensity signal to noise greater than 3. 
Peak brightness temperatures range from 1~K to 6.5~K.
The very evident bright \cii\ feature centered at (l,b) = (109.873$^\circ$, 4.267$^\circ$) or ($\alpha$,$\delta$) = 22$^h$47$^m$54.3$^s$, 63$^\circ$57'39.3''), is coincident with the peak of the extended dust dust emission seen at shorter ($\leq$ 250 $\mu$m) wavelengths \citep{Hennemann:2008, Ragan:2012}.  
This is distinct from the peak of the \co\ emission which, as seen in Figure~\ref{fig:fig1}, peaks at (l,b) = (109.85$^\circ$,4.26$^\circ$) or ($\alpha$,$\delta$) = (22$^h$47$^m$44.9$^s$, 63$^\circ$56'39.1'')
The remaining emission is fainter and more smoothly distributed within the cross--scan coverage.

At $\sim$1\arcmin\ angular resolution, the molecular gas distribution from CO appears centrally peaked 
on the 850\micron\ clump 
\citep{DiFrancesco:2008} and, as previously described, offset from the peak of \cii\ emission.  
The peak column density of molecular gas is 1.2$\times$10$^{22}$~\cmsq\ assuming a CO to \htwo\ conversion factor of 
2$\times$10$^{20}$~\cmsq(K\,km\,s$^{-1}$)$^{-1}$.  
The peak \htwo\ column density from the SCUBA dust emission is 6.5$\times$10$^{22}$ \cmsq.  
This value assumes a dust temperature of 12~K and a dust emissivity index of 2.  Given the uncertainties in both methods of column density determination, this is acceptably consistent

The \HI~21~cm line emission is displayed over the same area as the \cii\ and CO images with a color stretch beginning 
at 800 K km/s, which corresponds to a column density of 1.5$\times$10$^{21}$\cmsq\ assuming 
$N_H=1.823{\times}10^{18}\int T_Bdv$ \cmsq, where $T_B$ is the HI brightness temperature. 
The atomic gas is extended throughout the displayed field but shows a localized peak coincident with the CO emission maximum, 
with an atomic hydrogen column density of 2$\times$10$^{21}$ \cmsq.

The relative kinematics of each spectral line are accessible in the position--velocity images shown in the second and third columns of Figure~\ref{fig:fig1}. 
The brightest HI~21cm line emission is distributed over velocites $-49$ to $-43$ \kms\ with a peak at velocity $-45$~\kms.  
This peak velocity is blueshifted by $\sim$4~\kms\ relative to the peak velocity of \cii\ and CO emission of $-40.5$~\kms.  

\section{Tracing CNM with \cii\ emission}
The spatial and kinematic relationships between the atomic and molecular gas components are key factors in evaluating the role of converging flows of atomic gas in the formation of molecular clouds.
Given the total collisional velocity of 7.5~\kms\ in the simulations, \citet{Clark:2019} found that converging flows of cold, neutral atomic gas could be 
evident in \cii\ emission within  velocity intervals displaced several \kms\ from the central core of the line.  
However, they cautioned that the \cii\ emission from the CNM would be very faint, with antenna temperatures less than 0.1~K.

\begin{figure}
\centering
\includegraphics[width=0.65\columnwidth]{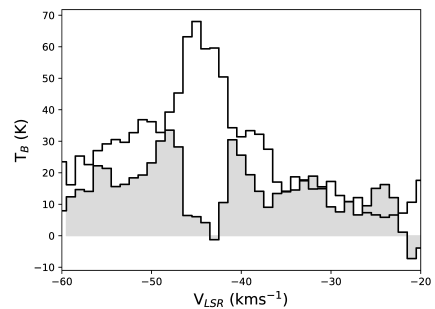}
\caption{\HI~21~cm spectra at the position of the extragalactic source G109.65+4.24 (filled in grey) and at the offset position at l=109.70$^\circ$, b=4.24$^\circ$. 
The absorption feature occurs over the velocity 
interval [$-48$,$-42$] \kms\ and 
indicates the presence of cold, neutral, atomic gas in the vicinity of the BKP~7323 molecular cloud. 
\label{fig:fig2}}
\end{figure}

Direct evidence for local CNM gas in the vicinity of the molecular cloud is provided by the absorption feature in the \HI~21~cm spectrum towards the extragalactic continuum source G109.65+4.24 located 0.2 degrees (11.3~pc projected offset) from the CO peak shown in Figure~\ref{fig:fig2} \citep{Strasser:2007}.   
The absorption occurs over the velocity interval $-48$ to $-42$~\kms.

\begin{figure*}
\centering
\includegraphics[width=0.8\textwidth]{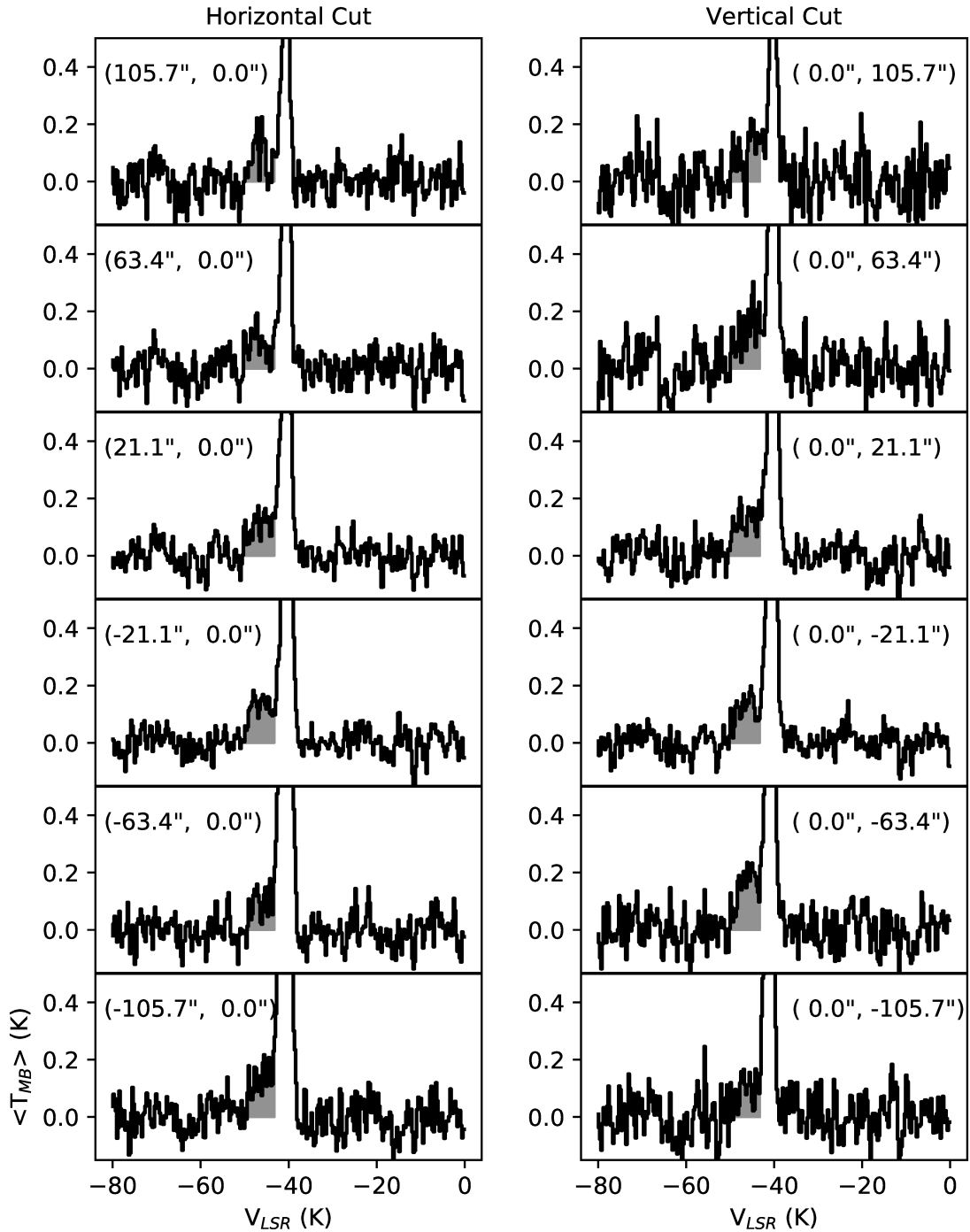}
\caption{\cii\ spectra averaged over 60 pixels along the horizontal (left) and vertical (right) cuts of the cross map.  
Blueshifted wing emission over the interval $-50$ to $-43$~\kms\ (filled in grey)  is evident. 
Angular offsets from the center of the map are shown in each spectrum. 
The measured brightness temperatures and velocity overlap with the absorption feature shown in Figure~\ref{fig:fig2} suggest that this \cii\ wing emission reflects CNM gas that is present throughout the BKP~7323 cloud. 
\label{fig:fig3}}
\end{figure*}

At its native resolution of 14.1\arcsec, the sensitivity of the \cii\ data presented in this study is insufficient to detect the predicted faint signal. 
To search for the signature of CNM gas associated with the cloud, we average the \cii\ spectra along the horizontal and vertical cuts in 6$\times$10 and 10$\times$6 pixel blocks, respectively. 
The average spectra are shown in Figure~\ref{fig:fig3}.  
We find weak but significant signal over the velocity interval $-50$ $<$ \vlsr $<$ $-43$~\kms\  that blends into the emission from the brighter \cii\ component centered at $-40.5$~\kms.   
The weaker component coincides with the \HI~21~cm absorption feature shown in Figure~\ref{fig:fig2} as well as the \HI~21~cm emission shown in the 
position velocity images in Figure~\ref{fig:fig1}.  
This component is detected in all averaged spectra, with the strongest  signal located near the center of the map, which provides circumstantial evidence that this feature is assocated with the molecular cloud.  
Warm neutral  atomic gas is expected to be characterized by extremely weak \cii\ emission due to low excitation, and would not be detected with our sensitivity. 
It is unlikely that the observed, faint \cii\ emission component arises from molecular gas with a low CO fractional abundance as it is kinematically displaced from the CO emission at $-40.5$~\kms. 

\begin{figure}
\centering
\includegraphics[width=0.8\columnwidth]{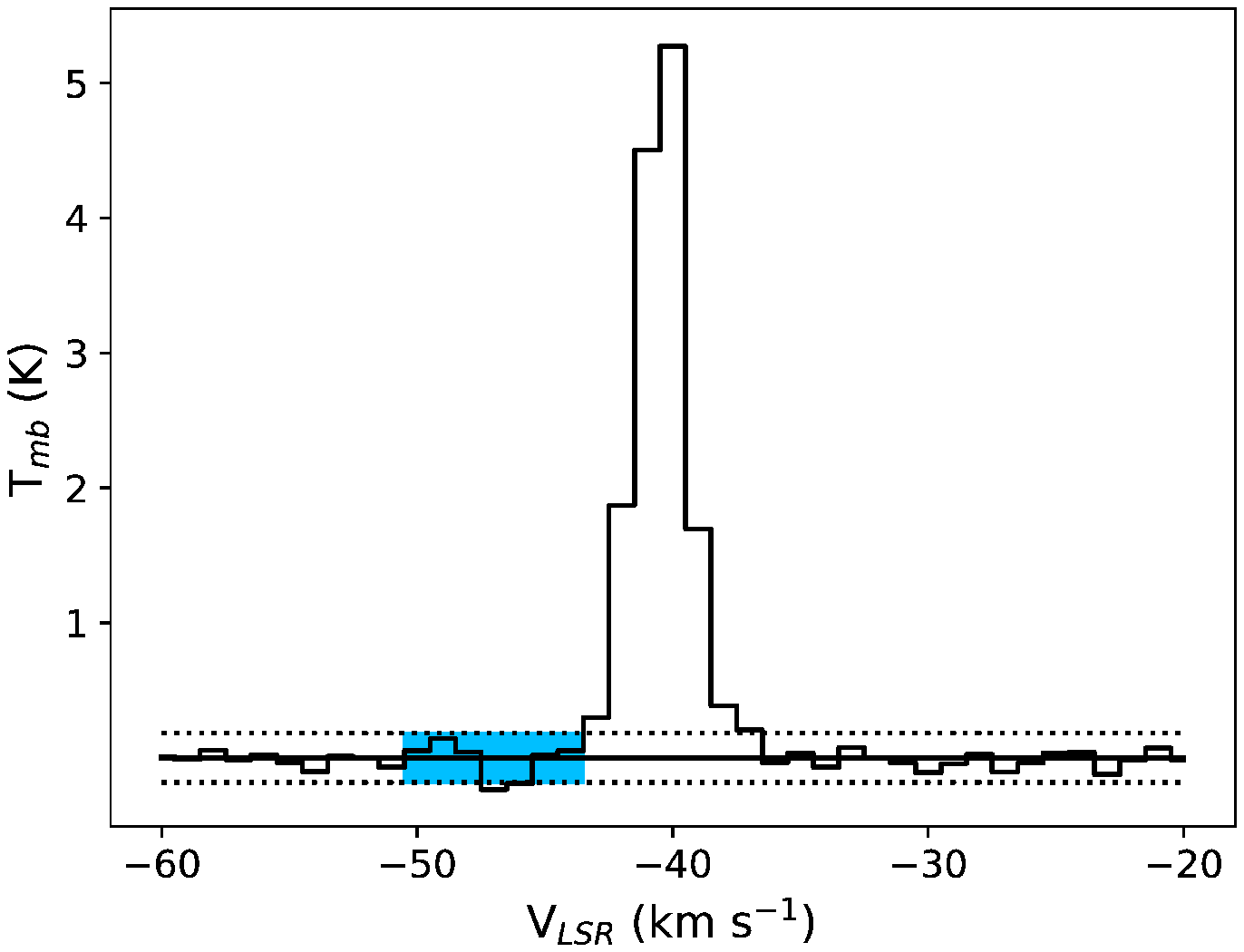}
\caption{\co\ spectrum averaged over the $0.1^\circ{\times}0.1^\circ$ field shown in Figure~\ref{fig:fig1}.  
The solid horizontal line marks T$_{mb}$=0 and the horizontal dotted lines 
mark T$_{mb}$=$\pm$3$\sigma$.  No \co\ emission 
is found within a 3$\sigma$ limit of 0.18~K over the velocity interval [$-50$,$-44$]~\kms\ (shaded in blue) that coincides with the 
blueshifted wing of \cii\ emission. 
\label{fig:fig4}}
\end{figure}

It is important to establish whether there is CO emission within the blueshifted wing interval of \cii\ emission. 
First, we assign the upper limit of the search interval to $-44$~\kms\ to avoid contributions from the wing of the CO main component at $-40.5$~\kms. 
No detected integrated CO emission is found in any of the spectra within the interval [$-50$,$-44$]~\kms\  above  the 3$\sigma$ limit of 
6.6~K\,\kms. 
Figure~\ref{fig:fig4} shows the CO spectrum averaged over the 
displayed field of Figure~\ref{fig:fig1}.   
No associated CO emission within the [$-50$,$-44$]~\kms\ interval is evident to a 3$\sigma$ limit of 0.18~K in antenna temperature and 0.40 K\,\kms\ in integrated emission.
Given the absence of CO emission in this interval, we connect this \cii\ emission to CNM atomic gas.  

To further quantify faint signal from ionized carbon, we fit a 2--component gaussian profile to the average \cii\ spectra displayed in Figure~\ref{fig:fig3} 
with Bayesian regression using the {\tt emcee} package \citep{Foreman-Mackey:2013}.  
The first gaussian component accounts for the bright \cii\ emission at $-40.5$~\kms\ associated with the photodissociation regions (see \S5)
while the second gaussian component describes the weak \cii\ emission at $-45$~\kms.

\begin{figure}
\centering
\includegraphics[width=0.8\columnwidth]{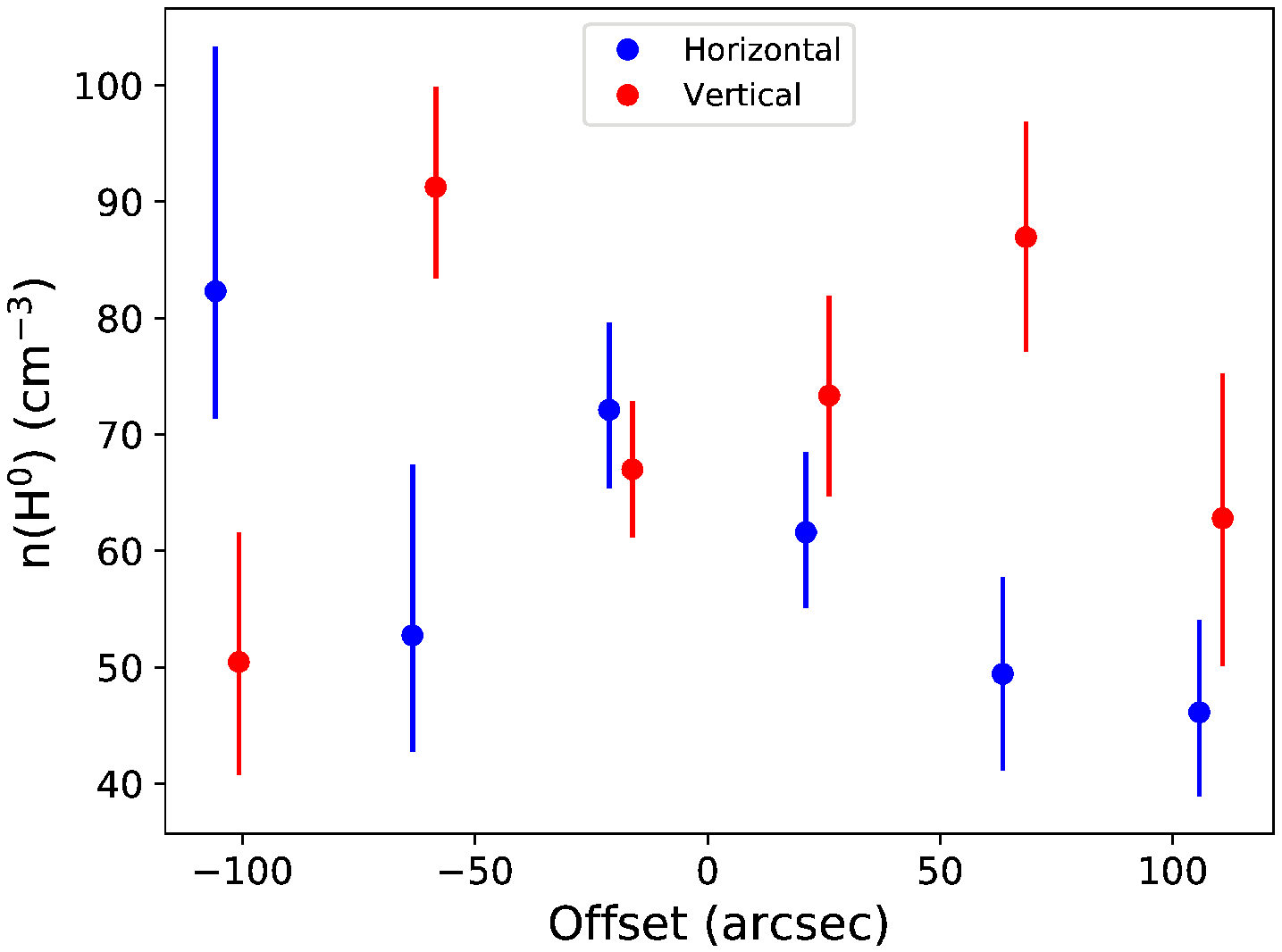}
\caption{Atomic hydrogen volume densities over the velocity interval [$-50$,$-43$] \kms\ derived from equation~\ref{eq:n0} for each average spectrum shown in Figure~\ref{fig:fig3}.  
The blue and red points correspond to average spectra long the horizontal and vertical cuts respectively.
\label{fig:fig5}}
\end{figure}

From the fitted  gaussian parameters of the wing emission, we derive the volume density of the atomic gas that is required to account 
for the observed \cii\ integrated intensity given the column density of ionized carbon.
Assuming the \cii\ emission comes exclusively from the atomic gas component in this velocity interval, the ionized carbon column density 
is estimated from the hydrogen column density derived from the \HI~21~cm line emission using
\begin{equation}
N(C^+)=[C/H] 1.823\times10^{18}\int T_B(v)dv~,
\label{eq:NC}
\end{equation}
where the integral is over the velocity interval $-50$ to $-43$~\kms\ and [C/H] is the carbon to hydrogen abundance ratio. 
A [C/H] value of 1.6$\times$10$^{-4}$ is adopted based on the results of \citet{Sofia:2004}.
We use equation 48 for the \cii\ fine structure line collisional deexcitation rate for collisions with atomic hydrogen from \citet{Goldsmith:2012} 
to derive the critical density of the [CII] line, n$_{cr}$ = 3030(Tk/100)$^{0.14}$ \cc. 
Then, using equation 2 of \citet{Langer:2010}, 
which relates integrated \cii\ intensities to column densities
assuming optically thin emission, we obtain 
\begin{equation}
n(H^0)=\frac{3.03\times10^3 (100/T_k)^{0.14} }{2 exp(-91.2/T_k)(3.43\times10^{-16}X-1)-1} \;\; cm^{-3}
\label{eq:n0}
\end{equation}
where $X=N(C^+)/W([CII])$.
$W$(\cii), is derived from the definite integral of the gaussian over all velocities, 
\begin{equation}
W([CII])=(2\pi)^{1/2} T_2\sigma_{v,2}~, 
\label{eq:gaussian_integral}
\end{equation}
and $T_2$ and $\sigma_{v,2}$ are the amplitude and velocity dispersion of the second gaussian component.  
In practice, we calculate a distribution of $n(H^0)$ values for each averaged spectrum by propagating values from the posterior distributions of $T_2$ 
and $\sigma_{v,2}$ through equation~\ref{eq:gaussian_integral} and equation~\ref{eq:n0}.  
A kinetic temperature of 100~K is assumed. 
The derived volume density profiles through the horizontal and vertical arms of the map (Figure \ref{fig:fig1} upper left) are shown in Figure~\ref{fig:fig5}.   
The mean density of all points is 67~\cc.  

The \cii\ brightness temperatures, derived hydrogen volume densities, and velocities that agree with that of the \HI~21~cm absorption point to CNM gas in the vicinity of the BKP~7323 molecular gas. 
This atomic gas is kinematically displaced from the CO emitting material by $-4$ \kms.  
In \S6, we discusss the possible origins of this component and its relationship to the molecular gas of BKP~7323.

\section{PDR Conditions in the Cloud}
The bright \cii\ lines from BKP~7323 and their alignment in velocity with the CO lines between $-43$ to $-38$~\kms\ point to  an origin from  photodissociated gas within the cloud. 
While the cloud is far removed from any strong sources of UV radiation, newborn stars within the cloud are likely responsible for the dissociating photons.
\citet{Ragan:2012} identify 7 protostellar cores from {\it Herschel PACS} imaging with luminosities ranging from 5 to 10 L$_\odot$.  
Six of these sources  reside within the dense submillimeter clump found by \citet{Hennemann:2008} while 1 source (ISOSSJ22478+6357 ID=7) is coincident with the peak of \cii\ emission and the peak position of mid infrared dust emission \citep{Ragan:2012}.
Recent high resolution imaging of the 1.1~mm dust continuum emission with {\it NOEMA} show 15 fragments within the clump that indicate a developing small cluster of protostars \citep{Beuther:2021}. 

The measured intensities $I_{CII}$, $I_{CO}$, and the ratio $I_{CII}/I_{CO}$ can provide contraints to the gas conditions within the PDR regions. 
We first convert the integrated intensity of each spectrum from units of K\,\kms\ to units of ergs s$^{-1}$ cm$^{-2}$ sr$^{-1}$ using the factor, ${2k_B}/\lambda^3$, where $k_B$ is the Boltzmann constant and $\lambda$ is the wavelength of the \cii\ fine structure transition. 
This conversion is appropriate for a uniform source filling a diffraction--limited beam. 
To evaluate the above quantities, we convolve and resample the \cii\ antenna temperatures to the angular resolution and sampling of the \co\ data. 

\begin{figure*}
\centering
\includegraphics[width=0.8\textwidth]{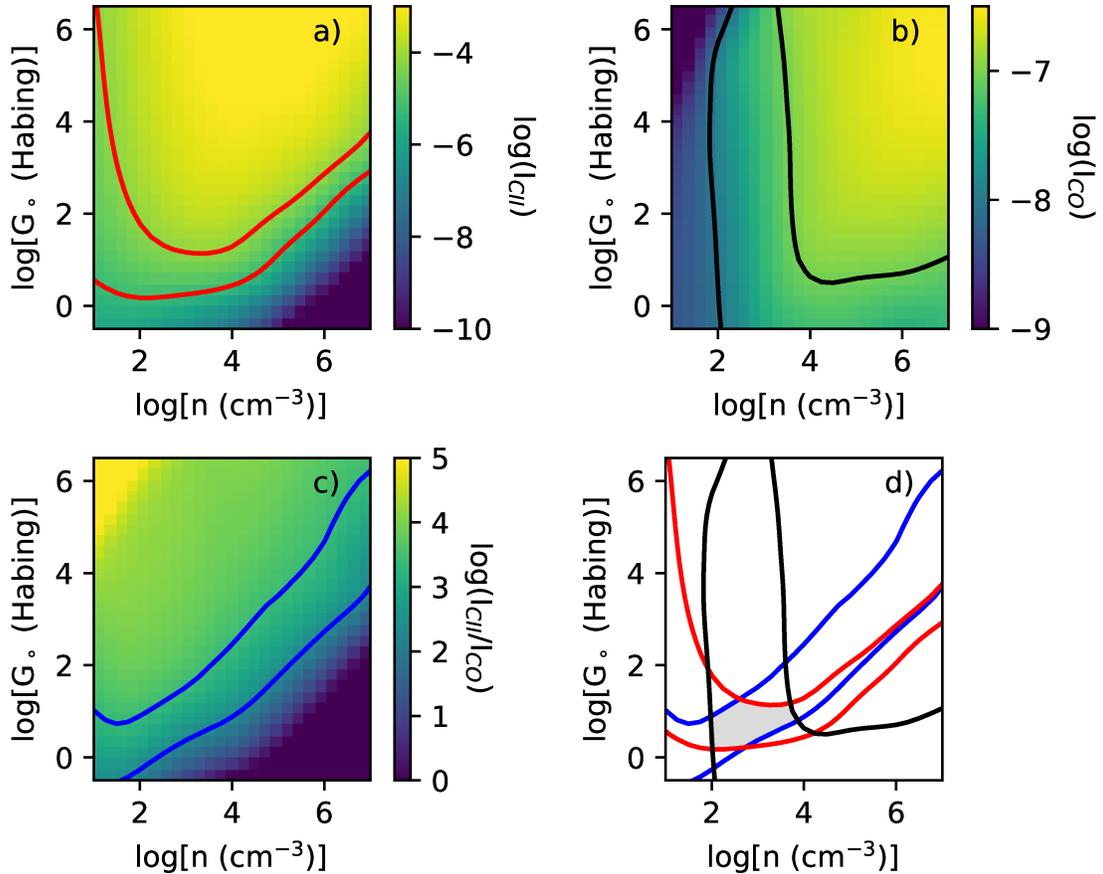}
\caption{{\it PDR Tool Box} models of line emission. a) Image of the logarithm of \cii\ intensities over a range of radiation fields and gas densities.
The red lines show contours  of the logarithms of the observed 3$\sigma$ detection limit ($-5.14$) 
and maximum \cii\ intensity ($-4.14$) (in units of ergs s$^{-1}$ cm$^{-2}$ sr$^{-1}$).
b) Same as a) for \co\ emission.  
Contours are the logarithm of the observed 3$\sigma$ detection limit ($-7.98$) and maximum intensity of \co\ emission ($-7.02$). 
c) Same as a) for the ratio  \cii/\co. 
Blue lines denote the logarithms of the 3$\sigma$ (2.42) and maximum (3.41) ratio values.
d) Same contours in a), b), and c).  
The grey shading marks subset area bounded by the contours of acceptable values of $I_{CII}$, $I_{CO}$, and $I_{CII}/I_{CO}$.  
This area provides the range of model radiation field and gas density values that are consistent with the observations.
\label{fig:fig6}}
\end{figure*}

The {\it PDR Tool Box} \citep{Pound:2008} compiles a set of PDR models for a range of interstellar radiation field 
intensities,  volume densities, and metallicities. 
The calculations include excitation and radiative transfer to produce model \cii\ and \co\ intensities and ratios. 
We examine model {\tt wk2006} from the {\it PDR Tool Box}, which assumes a metallicity Z=1 and spans the range of incident UV flux from 10$^{-0.5}$ to 10$^{6.5}$ Habings and cloud densities 10$^1$ to 10$^7$ \cc.  
The full set of model parameters are listed in Table~1 of \citet{Kaufman:1999}. 
The model \cii, \co\ intensities and $I_{CII}/I_{CO}$ values are shown as  images in Figure~\ref{fig:fig6} for a range of radiation field intensities, $G_\circ$ in Habing units and gas volume densities, $n_H$.  
The contours show the 3$\sigma$ detection limit and the maximum value for each line (6a,6b) and their ratio (6c) from our data.  
The shaded area in Figure~\ref{fig:fig6}d shows the range of log(G$_0$) and log($n_H$) where these measured extremes overlap.  
It covers the range of radiation field intensity 0.5 $<$ log(G$_0$) $<$ 1.5 and volume density range 10$^2$~\cc~ $<$ $n_H$ $<$ 10$^{4}$~\cc.
The range of allowable $G_\circ$ and $n$ is reasonable given the moderate luminosities from the developing protostellar cores identified by  \citet{Ragan:2012} and the volume densities probed by the  \co\ J=1--0 transition. 

\section{Discussion}
Our \cii\ observations have identified a faint emission component that is kinematically offset from the velocity of CO emission from the BKP~7323 molecular cloud. 
We attribute this component to CNM atomic gas based on the match between its 
velocity and that of a nearby \HI~21~cm absorption feature, the derived volume densities, and the measured \cii\ brightness temperatures that are similar to those predicted for CNM conditions. 
However, the physical relationship of this material to the molecular cloud is difficult to determine unambiguously. 
This component could simply be a distinct CNM cloud that has no physical connection to BKP~7323. 
Yet, the \HI~21~cm emission at this velocity has a local maximum at the position of the cloud, which is circumstantial evidence 
of a physical association. 
If related to the molecular cloud, it is unlikely that this component arises from a low column density, atomic envelope.  
To remain bound to the central, molecular cloud, the velocity of the envelope with respect to the CO velocity 
should be comparable to the virial velocity dispersion of the cloud. 
For a spherical cloud with uniform density and neglecting surface pressure 
and magnetic pressure terms, the virial velocity dispersion is 
(GM/5R)$^{1/2}$ \citep{McKee:1992}.  
The molecular mass of BKP~7323 is 1100~\msun\ and its radius is 3~pc.  So its 
virial velocity dispersion is 0.6~\kms, which is far 
less than the measured 4~\kms\ displacement of this component from the molecular gas velocity.  
Low density atomic gas that might be present in an envelope of the molecular cloud  thus would rapidly separate from the cloud with diameter $L$ 
within a crossing time, L/${\delta}$v$\sim$1.5~Myr.  

The absence of CO emission in this blueshifted interval (see Figure~\ref{fig:fig4}), could indicate that the \cii\ emission arises from a dark-CO component, where \htwo\ is present but 
CO emission is not detectable owing to very low CO abundance \citep{Wolfire:2010}. 
However, based on the numerical simulations by \citet{Clark:2019},  the velocity of a dark-CO component is expected to be comparable to the velocity of CO emission.  The 4~\kms\ 
velocity offset of the \cii\ blue shifted emission from the CO velocity indicates that the \cii\ emission is not from a CO-dark component of gas. 

\begin{figure*}
\centering
\includegraphics[width=0.8\textwidth]{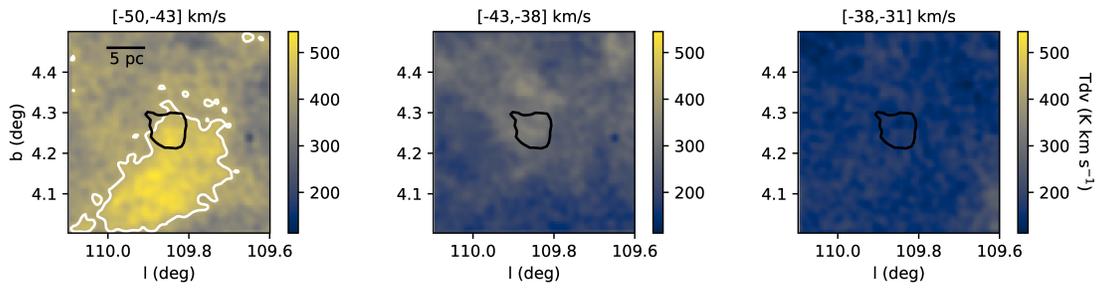}
\caption{Integrated HI~21~cm line emission over the velocity intervals (left) [$-50$,$-43$], (middle) [$-43$,$-38$], and (right) [$-38$,$-31$]~\kms. 
The black contour is the 5~K~km~s$^{-1}$ isophote of CO $J$ = 1 -- 0 emission.  The white contour in the [$-50$,$-43$]~\kms\ image corresponds to 
an atomic column density of 8$\times$10$^{20}$ \cmsq.
\label{fig:fig7}}
\end{figure*}

We propose that this emission component at \vlsr\ $\sim-$45~\kms\ represents a large scale flow of CNM material that is streaming onto the molecular cloud.  
But the question whether this flow is one part of a two stream, converging flow remains.
Ideally, a signature of converging flows onto a developing molecular cloud would be  a far--side, 
blue--shifted component and near--side, red--shifted component relative to the  bulk velocity of the molecular cloud 
as implied by the simulations by \citet{Clark:2019}.  
However, it is improbable that most converging flows 
are head--on collisions as modeled in the simulations. 
Our observations find no evidence for \cii\ emission over the velocity interval $-40$ to $-35$ \kms\ that would confirm the presence of a red--shifted 
counterpart to the observed blue--shifted gas flow indicated by the wing seen in Figure \ref{fig:fig3}.  
To examine the atomic gas directly over a larger area, we show  images of the \HI~21~cm line emission integrated over the \vlsr\ intervals [$-50$,$-43$], [$-43$,$-38$], and
[$-38$,$-31$]~\kms\ in Figure~\ref{fig:fig7}.  
The redshifted interval [$-38$,$-31$]~\kms\ shows faint HI emission with $W$(\HI) $<$ 200 K km s$^{-1}$ with no apparent connection to the CO cloud.  
This emission mostly likely represents diffuse WNM gas that is not interacting with BKP~7323. 

HI emission in the  [$-43$,$-38$]~\kms\ velocity range is spatially coincident  with the molecular cloud and extends to higher Galactic latitudes beyond the molecular 
cloud edge as traced by CO emission.  
All or part of this feature may be related to the atomic envelope of the molecular cloud.  
If a fraction of this atomic gas is streaming towards the molecular cloud as part of a counterpart, secondary flow, its motion would be in the plane of the sky.
However, such a streaming component would not be identified in \cii\ measurements as the bright emission from the PDR would greatly exceed the 
faint signal from the hypothetical CNM flow
within the same velocity interval. 
Alternatively,  the blueshifted component of CNM gas in the [$-50$,$-43$]~\kms\ interval  that is observed could be a singular flow onto the developing molecular cloud 
with no counterpart flow of atomic gas.
This component is brighter in \HI~21~cm line emission than the other components and provides a larger reservoir of material that could be incorporated into 
the molecular cloud. 

Taking the properties from the singular flow that we actually measure in the blue-shifted interval, 
the rate of atomic mass falling onto the cloud can be estimated as,
\begin{equation}
dM/dt = \rho v_{flow} L^2 = n(H^0) \mu_H m_H v_{flow} L^2~,
\end{equation}
where $\rho$ is the mass volume density, $v_{flow}$ is the flow velocity, and $L$ is 
the physical size of the cloud.  For $n(H^0)$=67~cm$^{-3}$, $v_{flow}$=4~\kms, and $L$=6~pc, 
$dM/dt$=3.2$\times$10$^{-4}$ M$_\odot$/yr.  At this mass flow rate, it would take 3.4~Myr 
to build up the current 1100 M$_\odot$ mass of the cloud.  

This time scale to form a molecular cloud or a clump within a larger cloud complex is a reasonable estimate 
with respect to computational simulations of converging flows \citep{Clark:2012,Clark:2019, Vazquez:2017}. 
In these studies, \htwo\ and CO molecules form within the 
dense regions generated by the shocks of the converging flows or within the ensuing thermal instabilities.  But this occurs in 
the later stages  -- 10-16~Myr after the initial interaction of the atomic gas flows as tracked by the simulations with low combined velocity collisions.  
This time scale 
is compatible with the 3.4~Myr formation time of BKP~7323. 

BKP~7323 is a small, low mass molecular cloud 
so its formation may not be representative of giant molecular clouds with masses greater than 10$^5$~\msun. 
Its spatial and kinematic isolation from other CO--emitting clouds precludes its having formed by
the agglomeration of smaller, existing molecular clouds.  Rather, the location of BKP~7323 in the outer Galaxy, where 
the molecular gas fraction is low, suggests
that it condensed directly from the neutral atomic gas component 
of the ISM \citep{Koda:2016}.  Our \cii\ measurements and \HI~21~cm data identify
a stream of cold, neutral atomic gas onto BKP~7323 that supports the top--down model of molecular cloud formation and growth in the outer Galaxy.

\section{Conclusions}
Observations of the \cii\ line from the compact molecular cloud BKP~7323 show bright emission that likely 
arises from the photodissociated gas resulting from embedded star formation activity within the cloud.  
Spatial averaging of the data reveals a weak secondary emission component that we attribute to cold, neutral atomic gas based on brightness
of the \cii, its having the same velocity interval as the nearby \HI~21~cm absorption feature, and the derived atomic gas mean volume density of 67~\cc.
The velocity of this component is blueshifted from the molecular cloud by $\sim$4~\kms, which suggests that this feature is  
atomic material flowing onto the molecular cloud with a mass infall rate of 3.2$\times$10$^{-4}$ {\msun}yr$^{-1}$.  Such atomic gas streams may 
provide an important reservoir of material to form molecular clouds and accumulate mass.

\acknowledgments
Our study is based in part on observations made with the NASA/DLR Stratospheric Observatory for Infrared Astronomy (SOFIA). 
SOFIA is jointly operated by the Universities Space Research Association, Inc. (USRA), under  NASA contract NNA17BF53C, and the Deutsches SOFIA Institut (DSI) under DLR contract 50 OK 0901 to the  University of Stuttgart. 
This work was supported by the Collaborative Research Centre 956 
funded by the Deutsche Forschungsgemeinschaft (DFG), project ID 
184018867 and by the Agence National de Recherche (ANR/France) and the 
Deutsche
Forschungsgemeinschaft (DFG) through the joint project “GENESIS” 
(ANR-16-CE92-0035-01/DFG1591/2-1).
Financial support for this work was provided by NASA through award 08-0062 issued by USRA.
This research was carried out in part at the Jet Propulsion Laboratory, which 
is operated by the California Institute of Technology under a contract 
with the National Aeronautics and Space Administration (80NM0018D0004).
This research used the facilities of the Canadian Astronomy Data Centre operated by the National Research Council of Canada with the support of the Canadian Space Agency.
We thank the SOFIA engineering and operations teams for their support that enabled the observations presented here. 
The authors also thank Nicola Schneider and Paul Casey for useful discussions and the referee for helpful comments.

\vspace{5mm}
\facilities{SOFIA (GREAT)}
\software{astropy \citep{astropy:2013}, emcee \citep{Foreman-Mackey:2013}, PDR Toolbox
\citep{Pound:2008,Kaufman:2006}}

\clearpage
\bibliography{bkp7323.bib}{}
\bibliographystyle{aasjournal}

\end{document}